\documentclass[sigconf,screen]{acmart}
\copyrightyear{2027}
\acmYear{2027}
\setcopyright{cc}
\setcctype{by}
\acmConference[SIGCSE TS 2027]{Proceedings of the 58th ACM Technical Symposium on Computer Science Education V.1}{February 17--20, 2027}{Sacramento, CA, USA}
\acmBooktitle{Proceedings of the 58th ACM Technical Symposium on Computer Science Education V.1 (SIGCSE TS 2027), February 17--20, 2027, Sacramento, CA, USA}
\acmDOI{10.1145/3857222.3860169}
\acmISBN{979-8-4007-3075-7/2027/02}

\usepackage[utf8]{inputenc}
\usepackage[T1]{fontenc}
\usepackage[nameinlink]{cleveref}
\usepackage{makecell}
\usepackage{xcolor,tikz}
\definecolor{stepblue}{HTML}{2F6FED}
\newcommand{\Step}[1]{%
    \tikz[baseline=(C.base)]\node[
    circle,
    fill=stepblue,
    text=white,
    font=\normalfont,
    inner sep=0.2pt,
    minimum size=2.2ex,   
    text width=2.2ex,   
    align=center
  ] (C) {#1};%
}

\begin{document}

\title{Rubric-Based Assessment of Student Reports Using~Small~Language Models}

\author{Sandesh Kumar Kafle}
\orcid{0009-0002-4275-1171}
\affiliation{
    \department{Faculty of Informatics}
    \institution{Masaryk University}
  \city{Brno}
  \country{Czech Republic}
} 
\email{xkafle@fi.muni.cz}

\author{Jan Vykopal}
\orcid{0000-0002-3425-0951}
\affiliation{
    \department{Faculty of Informatics}
    \institution{Masaryk University}
    \city{Brno}
    \country{Czech Republic}
}
\email{vykopal@fi.muni.cz}

\author{Valdemar Švábenský}
\orcid{0000-0001-8546-280X}
\affiliation{
    \department{Faculty of Informatics}
    \institution{Masaryk University}
    \city{Brno}
    \country{Czech Republic}
}
\email{valdemar@mail.muni.cz}

\renewcommand{\shortauthors}{Sandesh Kumar Kafle, Jan Vykopal, \& Valdemar Švábenský}

\begin{abstract}
Language models can analyze and assess student texts using predefined rubrics and deliver structured feedback, reducing instructors' workload. However, existing solutions often rely on external services that are resource-intensive, costly, and raise privacy concerns, making them unsuitable for constrained environments. This study evaluates an automated assessment pipeline with three key aims: design for low-resource environments, privacy, and focus on a novel context in cybersecurity education. The pipeline integrates lightweight language models for textual evaluation and a vision-language model for image interpretation. Responses from 105 students were collected over two semesters of an undergraduate course during a cybersecurity exercise on incident response. Results across different hardware configurations show that small models in low-resource environments can achieve up to 97\% accuracy in text-only and up to 94\% in rich-text assessments. This rivals or even outperforms a large model while adding the benefits of locally hosted assessment. Our solution can help educators and practitioners who need grading assistance without replacing human oversight, especially in environments where cloud-based solutions may be impractical. This study contributes an open-source, locally deployable AI-assisted scoring pipeline that runs on affordable hardware and can be extended to various types of student reports in computing education. As a result, it makes automated assessment accessible to a broad range of educators and institutions.
\end{abstract}



\maketitle

\section{Introduction} \label{introduction}

Generative AI (GenAI), especially Large Language Models (LLMs), is shifting learning environments from fully human-involved tutoring to automation in order to reduce instructors' workload. Classrooms are becoming more interactive~\cite{Yan2024}; LLMs are applied across learning contexts for short-answer grading~\cite{10.1007/978-3-031-11644-5_38,10.1007/978-3-031-36272-9_36}, essay summarization, automated feedback~\cite{10.1145/3641554.3701972,10.1145/3641554.3701801,10.1145/3706468.3706566}, and programming support~\cite{10.1145/3770762.3772539,10.1145/3770762.3772576}.

Nevertheless, existing studies do not address the context of rich-text reports, such as those created in cybersecurity exercises. In addition, computational accessibility remains a barrier to adoption. Many instructors lack access to high-end hardware or cloud-based infrastructures~\cite{Rodrigo2024,9835322,10746447}. While some institutions provide shared computing resources, there is a growing need for tools that can run effectively on moderate setups. 

Data privacy is also a concern for the education community~\cite{klose2020edm,ryan_s_baker_2024_12729816}. However, using cloud-based models or outsourcing computation to external services often means losing control of student data. In contrast, locally deployed models help preserve privacy, as there is no need to send sensitive data to a third party. Processing data locally reduces the risk of unauthorized access or misuse. 

This study evaluates the effectiveness of language models for text grading and feedback. It compares the performance and accuracy of five models across four hardware configurations, validating model output against human assessment. We analyze how small, open-source models perform in assessing student documents (both simple and rich-text) using a rubric -- particularly in low-resource setups prevalent in many educational contexts across the globe~\cite{Rodrigo2024}. Further, our work addresses the identified gap by providing a lightweight, locally deployable solution that maintains strong performance without requiring costly hardware~\cite{supplementary-materials}. 

Unlike open-ended questions and essay grading, we focus on structured student reports, aiming not only to interpret meaning or assess argumentation but also to accurately parse and evaluate predefined formatting, embedded content, and uniformity across submissions. In doing so, this study explores the potential of language models to save instructors' time for evaluation and feedback.

\section{Related Work} \label{relatedwork}

\subsection{Automated Assessment Challenges}

Scaling automated assessment is challenging, particularly when providing real-time, context-aware feedback to many students~\cite{10.1007/978-3-031-98417-4_23,ZamfirescuPereira2025}. Running large, state-of-the-art models requires specialized GPU clusters, enough memory, and disk space. If such resources are unavailable, using smaller, more computationally accessible variants is needed~\cite{Zhang2025,preprint-wei2025smallsignificantpromisesmall}. Furthermore, privacy regulations such as FERPA and GDPR must be considered, as transmitting student data to third parties can lead to information leakage or ethical issues~\cite{preprint-wang2025,KEHKASHAN2025100793}. To address these concerns, institutions have explored local deployment of open models, which can reduce costs and keep student data under institutional control~\cite{policar2025automated}. However, instructors may face system failures caused by resource overuse during peak submission periods, which can disrupt grading workflows~\cite{preprint-wang2025}.

\subsection{Efficient Language Models for Education}

Small models, such as \texttt{phi-3:3.8b}, \texttt{mistral:7b}, and \texttt{llama-3.2:3b}, have expanded the possibilities of AI-assisted education in limited environments~\cite{Zhang2025,preprint-wei2025smallsignificantpromisesmall}. These models can be deployed in resource-constrained settings with solid performance, reducing the need for larger models. CyberLLaMA, a fine-tuned \texttt{llama-3.2:3b} model for recognizing cybersecurity entities (e.g., malware names and IP addresses), is an example~\cite{Zhang2025}. Instruction-tuned models are increasingly used for specialized tasks such as reflection assessment and multi-session dialogue understanding~\cite{Li2025}. Prior work showed that students can receive help in mathematics, supporting the broader applicability of compact and instruction-tuned models beyond text generation~\cite{10.1007/978-3-031-98417-4_16}. Despite their computational advantages, smaller models can be less effective on complex prompts, requiring fine-tuning or careful task decomposition. For instance, TinyLlama-based study highlighted both the promise and limitations of compact models, depending on task complexity~\cite{preprint-zhang2024tinyllama}.

\subsection{Rubric-Aligned Assessment Methodologies}

Rubric-structured direct prompting can automate complex evaluations, in which language models follow expert-defined criteria and perform assessment~\cite{10.1145/3770762.3772539,policar2025automated,10.1007/978-3-031-98459-4_1}. These systems use forced-choice or ordinal scoring. Models respond with criterion-level results (e.g., correctness of a reasoning step, presence of a required element) and assign discrete values (e.g., 0/1/2) to reduce subjectivity and improve inter-rater consistency~\cite{Li2021}. To enable integration into learning platforms, outputs are constrained to structured formats (e.g., JSON) that include criterion scores and feedback~\cite{policar2025automated}. Few-shot grading systems, combined with guided rubrics and graded examples, enhance the accuracy and quality of grading for open-ended responses~\cite{10.1007/978-3-031-98459-4_1}. However, they are often used for grading essays or short answers rather than long, rich-text reports (such as cybersecurity incident reports, summaries, and advisories).

\subsection{Contribution of This Paper}

Prior studies on assessment primarily focused on structured short-answer evaluation, essay grading, and feedback generation in programming education using rubric-guided prompting~\cite{10.1145/3770762.3772576}. Next, deployment of automated assessment often assumed access to LLMs or did not consider low-resource contexts as a constraint \cite{Li2021,policar2025automated}. Our study addresses this gap by investigating smaller, efficient models for rubric-aligned scoring and feedback under constrained conditions. The proposed tool performs a rubric-based assessment of reports with rich formatting and contexts that may include both textual and visual evidence. Also, it speeds up the evaluation process, enabling students to receive an assessment report much more quickly, which supports their learning and helps them improve more effectively. We release the tool as open-source software~\cite{supplementary-materials}.

\section{Teaching Context} \label{educational_context}

The student reports analyzed in our study come from a cybersecurity simulation exercise held in two semesters (in 2024 and 2025) of an introductory undergraduate course, ``Cybersecurity in an Organization,'' taught at a Czech public university. The exercise was a real-time, in-person simulation of a cybersecurity incident taking place at a fictitious University. Students roleplayed as incident handlers on the University’s Computer Security Incident Response Team (CSIRT-U). Towards the end of the exercise, each student team was asked to prepare two text documents: an \textit{executive summary} and a \textit{security advisory}. The teams created the reports as plain text and PDF files, which were manually scored by exercise instructors according to assessment rubrics (see \Cref{sec:metrics}). The following text details the content and structure of both documents and rubrics.
 
\subsection{Student Report 1: Executive Summary (ES)} 

This document reports essential information about a phishing incident to the CSIRT-U's management. It summarizes key developments and findings, reporting on how the incident was handled. In particular, it provides information on incident reports received, how users who visited the phishing website were contacted, and whether their passwords were reset. It also explains how the phishing site was disabled and whether a security advisory was issued. \Cref{fig:es} depicts an exemplary ES created by the instructors. 

\textbf{\textit{Assessment Rubrics.}} \label{rubrics:es}
The rubric includes four criteria, with 4 points awarded by the grader if the report contains information: 
\begin{itemize}
    \item about incident reports CSIRT received (from an external party and an employee) -- 1~pt.
    \item that CSIRT contacted the users who visited their phishing website, and reset their passwords -- 1~pt.
    \item that CSIRT disabled the site hosting the phishing form -- 1~pt.
    \item that CSIRT issued a security advisory -- 1~pt.
\end{itemize}

\begin{figure}[ht]
\centering    
\scriptsize
\fbox{
\begin{minipage}{\columnwidth}
CSIRT-U received an incident report from an email provider, FREEMAIL, concerning phishing emails linking to one of the university's websites. This report was further complemented by one more report of phishing, now from our constituency, leading to the same university website.\\[-1.2mm]

CSIRT-U responded to this incident by tracing and contacting two users affected by the phishing campaign and requesting a password reset. CSIRT-U also contacted the website's hosting provider to temporarily turn off the website to prevent more users from becoming victims. The affected website is being returned to a secure state by its administrator.\\[-1.2mm]

A security advisory has been issued to all members of the University to lessen the impact of a potential second round of the campaign. Most importantly, the advisory includes a description of the incident, pictures of the phishing email and website, and user recommendations. 
\end{minipage}
}
\caption{Exemplary executive summary of the incident.}
\Description{Exemplary executive summary of the incident.}
\label{fig:es}
\end{figure}

\subsection{Student Report 2: Security Advisory (SA)}

\begin{figure}[!htbp]
    \centering
    \includegraphics[width=0.95\linewidth]{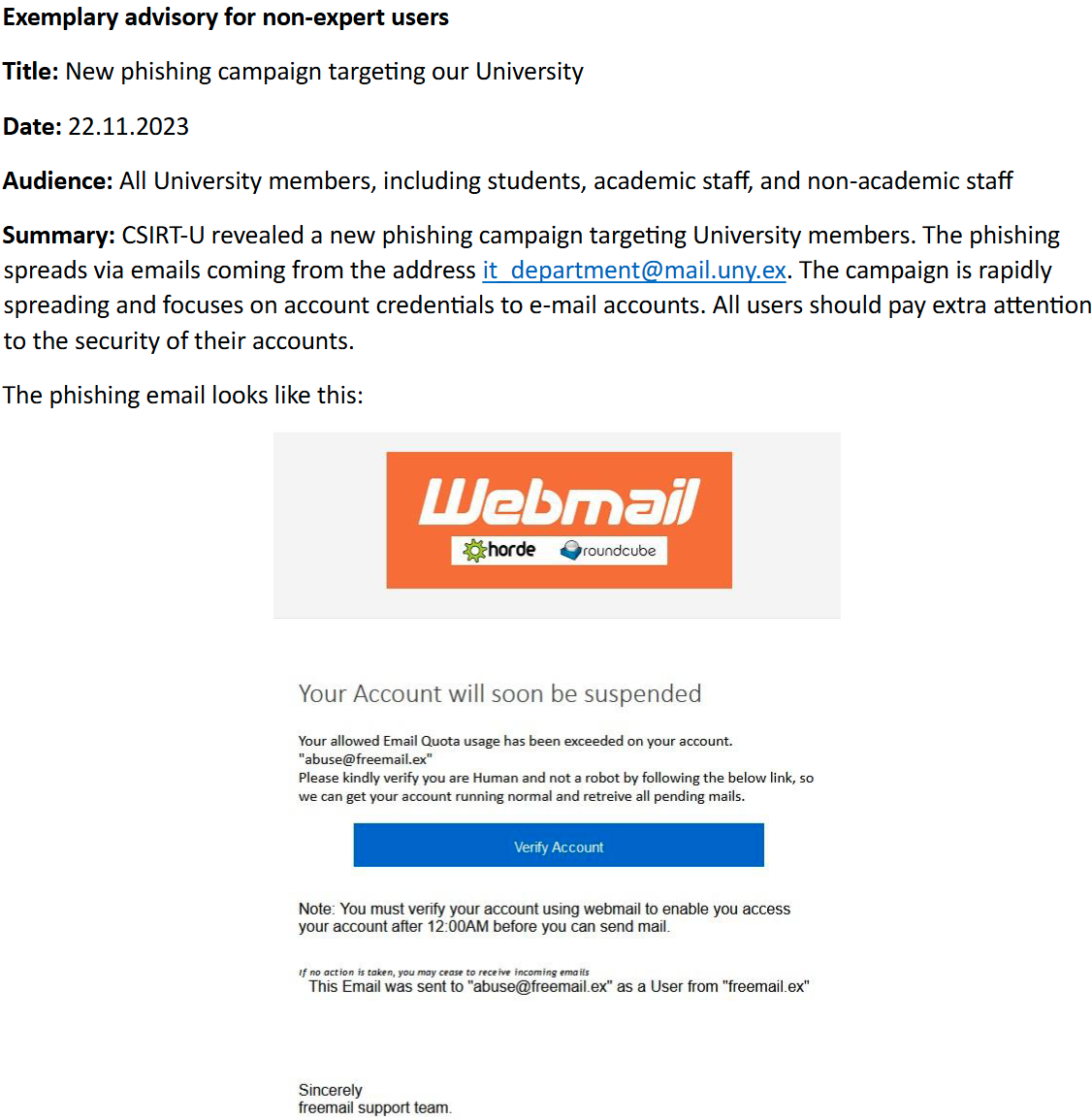}
    \includegraphics[width=0.95\linewidth]{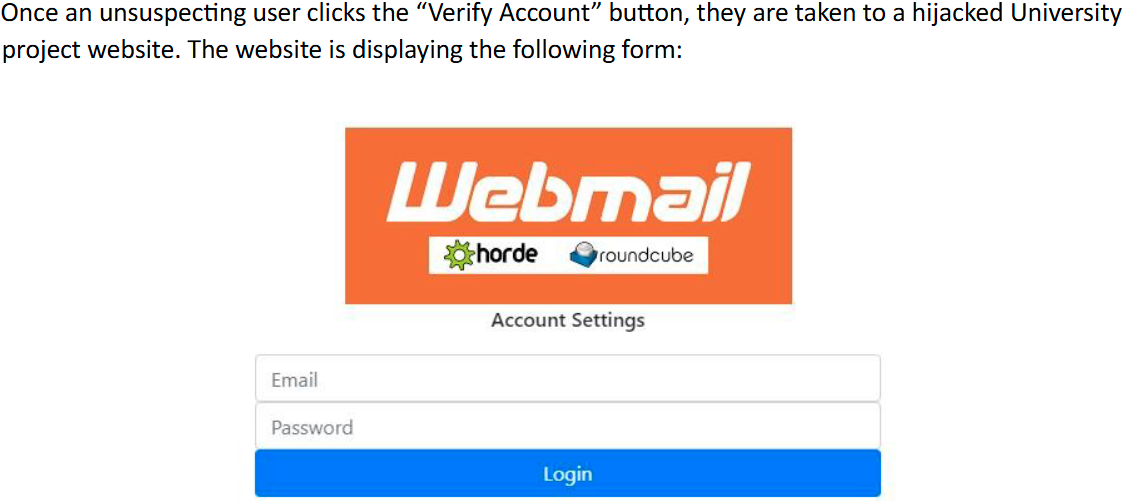}
    \includegraphics[width=0.95\linewidth]{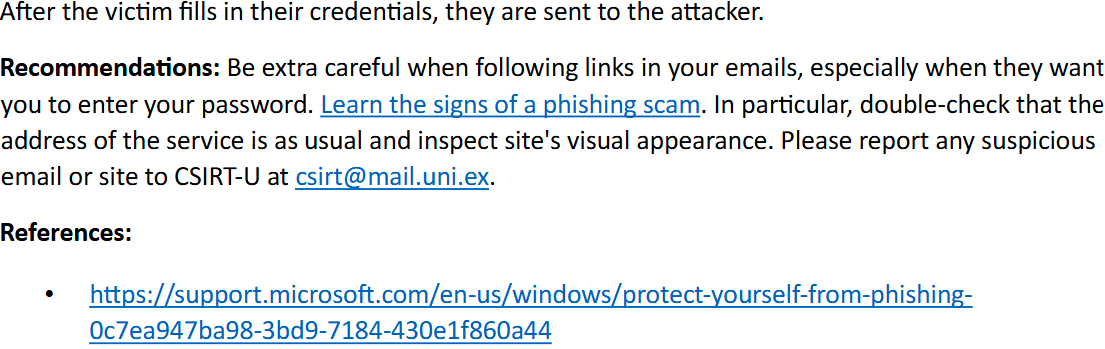}
    \caption{Exemplary security advisory.}
    \Description{Exemplary security advisory.}
    \label{fig:sa}
\end{figure}

In the exercise, students were tasked with informing University users about the ongoing incident via a security advisory. The SA should summarize what happened, present the phishing email and the corresponding malicious website, and provide recommendations for users with additional references. \Cref{fig:sa} depicts an exemplary SA, which must be a PDF document with two screenshots: a phishing email users may receive and the phishing website with a form, making it easier for recipients to recognize these threats. The expected structure of sections includes a title, date, target audience, summary, recommendations, and references. 

\textbf{\textit{Assessment Rubrics.}}\label{rubrics:sa}
The rubric includes four criteria, with a total of 5 points that can be awarded by the grader: 
\begin{itemize}
    \item Presence of all important sections (see above) and their meaningful content -- 2 pts when all good, 1 pt when one section is missing or has low quality, 0 pts otherwise.
    \item Screenshot of the phishing email -- 1 pt.
    \item Screenshot of the phishing website -- 1 pt.
    \item Inclusion of the phishing email sender address -- 1 pt.
\end{itemize}

\section{Research Methods for Report Assessment} \label{methods}

\begin{figure*}[ht]
    \centering
    \includegraphics[width=\linewidth]{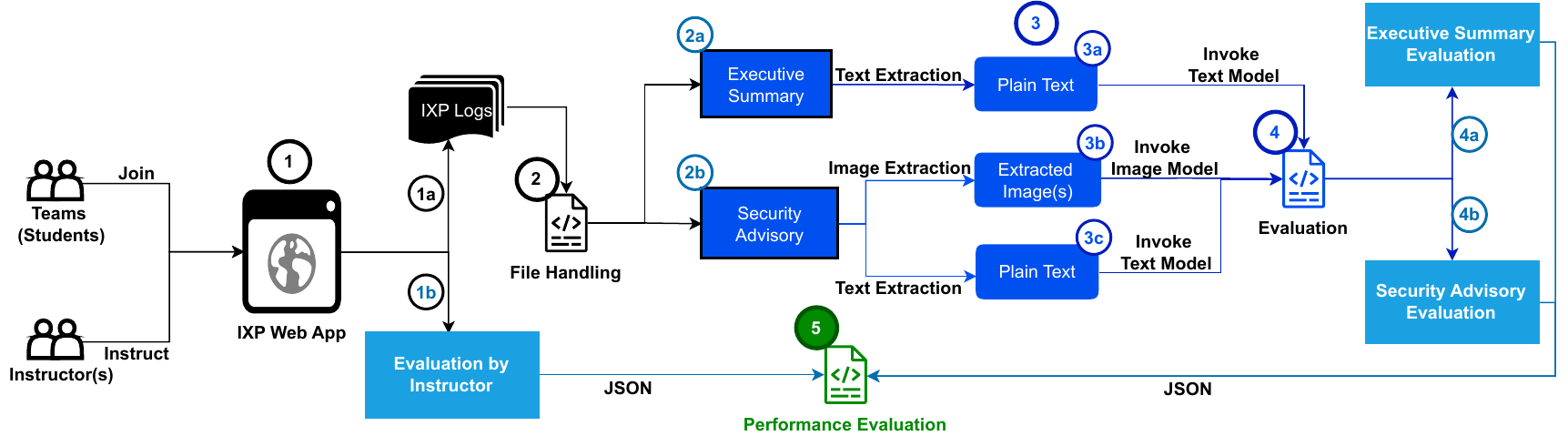}
    \caption{Illustration of the data processing pipeline and methods.}
    \Description{Illustration of the data processing pipeline and methods.}
    \label{fig:method}
\end{figure*}

While LLMs are often used for natural language processing~\cite{preprint-wei2025smallsignificantpromisesmall}, we investigated smaller models for automated assessment in low-resource, local deployment contexts. This approach aims to reduce resource consumption and make GenAI more accessible for assessment. \Cref{fig:method} outlines our data processing methods. The following subsections refer to steps marked as circled numbers in \Cref{fig:method}.

\subsection{Data Collection and Preprocessing} \label{data_collection}

This study did not require approval from the Czech university's institutional review board since only anonymized student texts were processed. We included only data from students who provided their informed consent to use their exercise data for research.

The exercise was conducted on the freely available INJECT Exercise Platform (further IXP)~\cite{10.1145/3649217.3653639}~(step \Step{1}). Out of 38 teams of three, 35 provided consent for this research. From the exercise logs (\Step{1a}), we analyzed two reports created by each team: the executive summary (\Step{2a}) and the security advisory (\Step{2b}). To clean the dataset, we removed duplicate submissions (e.g., cases where a team uploaded the same report multiple times) and kept only the final version for each team and report type (step \Step{2}). This step prepared the model's input data: email text and PDFs submitted by the students to the IXP. A Python script with widely used libraries \texttt{PyPDF2} and \texttt{Pillow} was employed to extract text and embedded images separately (step \Step{3}). The summary consisted of text only (\Step{3a}), whereas the text and images were handled separately as the advisory included both types of content (\Step{3b} and \Step{3c}). Since small models may lack the capacity to handle raw and noisy inputs on limited hardware, pre-processed and cleaned inputs help the model process them accurately. With minimal modifications, the entire processing pipeline can be adapted to handle a wide range of document types and structures, making it suitable for diverse educational contexts. 

\subsection{Data Analysis} \label{data_analysis}

Evaluating executive summaries (step \Step{4a}) is straightforward since they contain only text. In contrast, security advisories mix text and images, necessitating two models: one for text and another for images, because the text models cannot process images. We first used an image model to generate textual descriptions of the images (step \Step{3b}) while verifying their inclusion and correctness as required by the rubric. These descriptions were then combined with the extracted text from the advisory (step \Step{3c}) and evaluated with the text model (step \Step{4b}). This approach allows us to use a smaller model, reducing computational load. 

The analysis (step \Step{4}) was implemented in Python using the \texttt{langchain\_ollama} package. The models were hosted locally using Ollama~\cite{ollama}, an open-source platform for running language models on-device. Different hardware configurations with commonly available NVIDIA GPUs were used (see \Cref{tab:hw}). S1 uses a low-cost card from 2018, S2 is equipped with a new, yet low-power, laptop GPU, and S3 has an affordable desktop GPU. 
In addition, we tested a system without a GPU. The text-only evaluation took an average of 30 minutes per model. However, the CPU-only setup could not complete the rich-text evaluation. Even with a smaller image model, the output was incomplete, so we excluded it from our comparisons. 

The models were selected based on the hardware constraints (see~\Cref{tab:hw}). We used \texttt{llama3:8b} and \texttt{mistral:7b}, which follow instructions strictly and efficiently~\cite{10.1145/3641554.3701917}, enabling real-time response generation. We also used \texttt{qwen3:1.7b} and \texttt{gemma3:4b}, whose cost-effective reasoning makes them well-suited for complex tasks in resource-constrained environments~\cite{10.1145/3641554.3701917}. For image analysis, we used \texttt{llava:7b}, which generates textual descriptions of images. Prior work shows that \texttt{llava:7b} offers multimodal reasoning and delivers competitive accuracy across benchmarks, adapting well to both general-purpose and domain-specific image analysis tasks~\cite{10030474,10.1145/3735648}. 
Finally, to contextualize the possibilities of using small models on consumer-grade hardware, we also tested a larger \texttt{gpt-oss-120b} model, which achieves near-parity with Open\-AI o4-mini~\cite{openai2025gptoss120b}. This model was hosted on an on‑premise AI platform with one NVIDIA H100 GPU dedicated to this model. This remote system (denoted SR) serves only as a baseline for comparison with the locally hosted models.

\begin{table}[t]
\footnotesize
\centering
\caption{Hardware configurations used in our study.}
\label{tab:hw}
\setlength{\tabcolsep}{4pt} 
\begin{tabular}{|l|l|l|l|l|}
\hline
\textbf{System} & \textbf{GPU} & \textbf{VRAM} & \textbf{CPU} & \textbf{RAM} \\
\hline
    System 1 \textbf{(S1)} & RTX 2080Ti    & 12 GB & AMD EPYC 7F52         & 30 GB \\
    System 2 \textbf{(S2)} & RTX 5070Ti    & 12 GB & AMD Ryzen AI 9 XH 370 & 32 GB \\
    System 3 \textbf{(S3)} & RTX PRO 2000  & 16 GB & Intel i7-11700        & 64 GB \\
    \hline
    System 4 \textbf{(SR)} & H100 NVL      & 94 GB & 2$\times$ AMD EPYC 9454 & 1.5 TB \\
\hline
\end{tabular}
\end{table}

\subsection{Rubric-Aligned Prompt Design}

\begin{figure*}[t]
    \centering
    \includegraphics[width=\linewidth]{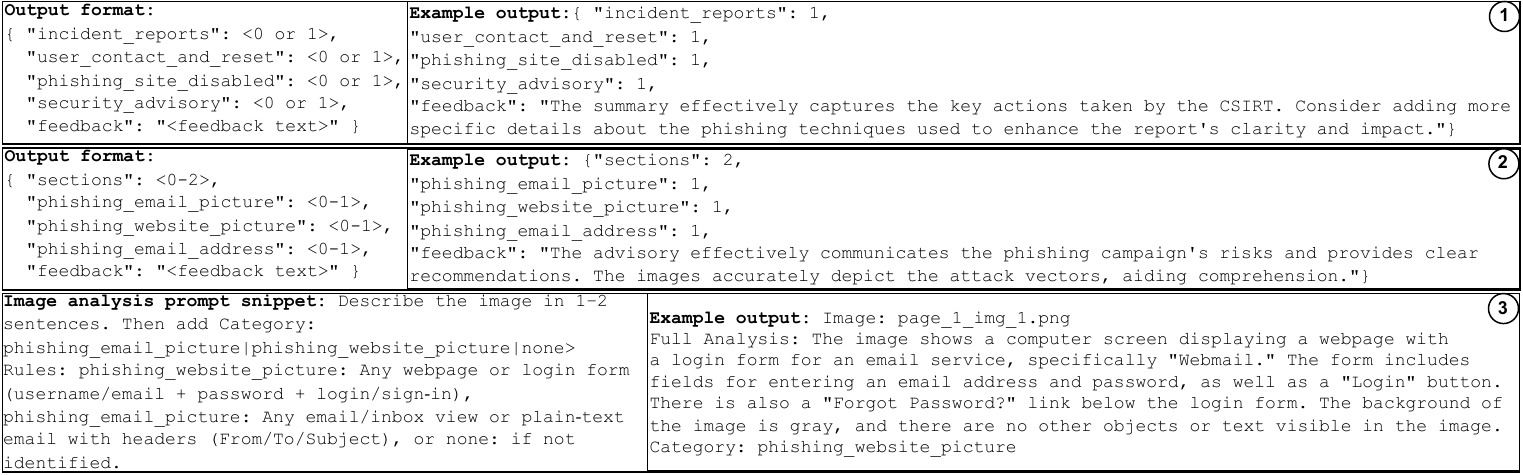}
    \caption{Defined output format and the generated output for (1) ES and (2) SA. And, (3) image analysis prompt and output.}
    \Description{Defined output format and the generated output, image analysis prompt and output.}
    \label{fig:OP_Examples}
\end{figure*}

We tried less-structured prompt techniques (e.g., zero-shot, one-shot, context-free), and response times were quick, but accuracy was not satisfactory. Here, we present the prompts developed after multiple test runs for assessing the student reports (described in \Cref{educational_context}). The prompts are designed for summative scoring and feedback, using language models to act as an instructor for evaluation. The model assesses student responses using a detailed rubric, then returns scores with feedback in JSON format. The prompts combine role assignment, context, assessment rubrics, an exemplary report, and output constraints (see details in~\cite{supplementary-materials}). Prompt templates are listed below; example output is in \Cref{fig:OP_Examples}.

\begin{footnotesize}
\begin{description}
\item[\textit{Executive Summary (ES) Assessment Prompt}:]
\item[Role:] You are an instructor for the university course Cybersecurity in an Organization, evaluating a student's executive summary following a simulated security incident for CSIRT-U.
\item[Context:] The student must write a concise executive summary (700–1000 characters) for university management, describing CSIRT's response actions, suitable for professional communication.
\item[Rubrics:] [copied from \Cref{rubrics:es}]
\item[Scoring instructions:]  Assess the student's executive summary. Evaluate strictly using the rubric, awarding 0 or 1 point for each criterion; provide only brief, constructive feedback and respond in the specified JSON format.
\end{description}

\begin{description}    
\item[\textit{Security Advisory (SA) Assessment Prompt}:]
\item[Role:] You are an instructor for the university course Cybersecurity in an Organization, evaluating a student's security advisory following a simulated security incident for the CSIRT-U.
\item[Context:] The student is tasked with writing a concise security advisory about an ongoing phishing campaign, addressed to all university staff and students. The advisory must include all critical information, two specific images (of the phishing email and phishing website), and be suitable for rapid dissemination within the university.
\item[Rubrics:] [copied from \Cref{rubrics:sa}]
\item[Scoring instructions:] Assess the student's security advisory and review the image analysis. Evaluate strictly using the rubric. Provide only brief, constructive feedback and respond in the specified JSON format.
\end{description}
\end{footnotesize}

\subsection{Metrics for Performance Evaluation} \label{sec:metrics}

Each student's report had from 56 to 305 words. Across the two exercise sessions, five instructors (2024) and six instructors (2025) graded different subsets of submissions. The grades were then checked and possibly updated by the main instructor (the course guarantor). The instructors also carefully reviewed all reports to avoid potential prompt injection by students; no such attempts were present. To measure the performance (step \Step{5}), we used \emph{evaluation accuracy} and \emph{processing time}, computed as:

\begin{small}
\begin{minipage}[t]{0.48\linewidth}
\centering
\[
\begin{aligned}
\text{Accuracy} &=
\frac{1}{N}\sum_{i=1}^{N}
\begin{cases}
1 & \text{if } m_i = h_i,\\
0 & \text{if } m_i \neq h_i
\end{cases}
\end{aligned}
\]
\end{minipage}
\hfill
\begin{minipage}[t]{0.5\linewidth}
\centering
\[
\begin{aligned}
\text{Processing time (avg.)} &=
\frac{1}{N}\sum_{i=1}^N t_i
\end{aligned}
\]
\end{minipage}
\end{small}

Where \(N\) is the total number of evaluated submissions, \(m_i\) is the model score for submission \(i\), \(h_i\) is the human score for submission \(i\), and \(t_i\) is the time (in seconds) taken by the model to evaluate submission \(i\). For each model across the systems, we evaluated each student report 10 times and measured the mean accuracy and mean evaluation time across these runs (see~\Cref{fig:sa-es-comparison}). This approach mitigates potential variation that could affect a single run.

\section{Results and Discussion} \label{results}

\begin{figure}[ht]
    \centering
    \includegraphics[width=1.00\linewidth]{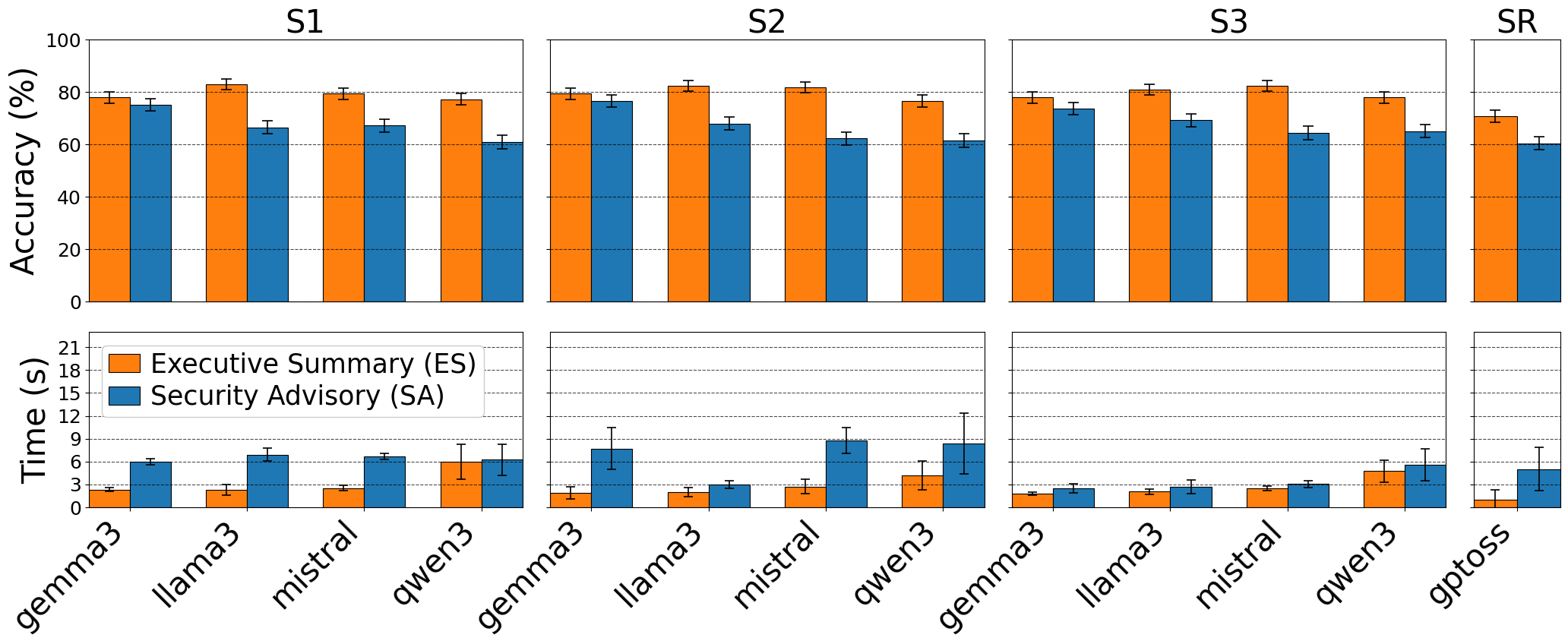}
    \caption{Results from 10 independent runs. 
    Overall accuracy (match \%) with 95\% CIs. 
    Evaluation time (s) with $\pm$ SD.}
    \Description{Results from 10 runs, which show overall accuracy with 90 \% CIs, evaluation time in seconds with SD.}
    \label{fig:sa-es-comparison}
\end{figure}

For each report type (ES and SA), the metrics (accuracy and time) are the mean across all teams. \Cref{fig:sa-es-comparison} presents the metrics when evaluating a single student report across the four systems (see \Cref{tab:hw}), with 95\% confidence intervals (CIs). \Cref{tab:accuracy_es} and \Cref{tab:accuracy_sa} report the per-criterion accuracy of evaluating ES and SA, respectively. 

\begin{table}[t]
\scriptsize
\caption{Per-criteria Evaluation Accuracy (\%) for ES (See~\Cref{rubrics:es} and~\Cref{sec:metrics}) for a single run.}
\label{tab:accuracy_es}
\setlength{\tabcolsep}{2pt}
\centering
\begin{tabular}{|l|c|c|c|c|c|c|c|c|c|c|c|c|}
\hline
\multicolumn{1}{|r}{\textbf{Criteria}} &
\multicolumn{3}{|c|}{\makecell{\texttt{incident}\\\texttt{\_reports}}} & 
\multicolumn{3}{c}{\makecell{\texttt{phishing\_site}\\\texttt{\_disabled}}} &
\multicolumn{3}{|c|}{\makecell{\texttt{security}\\\texttt{\_advisory}}} &
\multicolumn{3}{c|}{\makecell{\texttt{user\_contact}\\\texttt{\_and\_reset}}} \\ 
\cline{1-13}
\textbf{Model} 
& S1 & S2 & S3  
& S1 & S2 & S3  
& S1 & S2 & S3  
& S1 & S2 & S3 \\
\hline
\texttt{gemma3}  & 68.57 & 68.57 & 68.57  & 97.14 & 97.14 & 97.14  & 68.57 & 68.57 & 68.57  & 77.14 & 77.14 & 77.14 \\
\texttt{llama3}  & 68.57 & 68.57 & 68.57  & 91.43 & 91.43 & 88.57  & 85.71 & 85.71 & 85.71  & 85.71 & 82.71 & 80.00 \\
\texttt{mistral} & 68.57 & 71.43 & 71.43  & 85.71 & 85.71 & 88.57  & 77.14 & 82.86 & 80.00  & 85.71 & 85.71 & 88.57 \\
\texttt{qwen3}   & 71.43 & 68.57 & 71.43  & 97.14 & 94.29 & 88.57  & 65.71 & 65.71 & 71.43  & 74.29 & 77.14 & 80.00 \\
\hline
\texttt{gptoss} & \multicolumn{3}{c|}{45.71} & \multicolumn{3}{c|}{88.57} & \multicolumn{3}{c|}{82.86} & \multicolumn{3}{c|}{57.14} \\
\hline
\end{tabular}
\end{table}

\begin{table}[t]
\caption{Per-criteria Evaluation Accuracy (\%) for SA (See~\Cref{rubrics:sa} and~\Cref{sec:metrics}) for a single run.}
\scriptsize
\label{tab:accuracy_sa}
\setlength{\tabcolsep}{2pt}
\centering
\begin{tabular}{|l|c|c|c|c|c|c|c|c|c|c|c|c|}
\hline
\multicolumn{1}{|r}{\textbf{Criteria}} &
\multicolumn{3}{|c}{\makecell{\texttt{phishing\_email}\\\texttt{\_address}}} & 
\multicolumn{3}{|c}{\makecell{\texttt{phishing\_email}\\\texttt{\_picture}}} &
\multicolumn{3}{|c|}{\makecell{\texttt{phishing\_}\\\texttt{website\_picture}}} & 
\multicolumn{3}{c|}{\texttt{sections}} \\ 
\cline{1-13}
\textbf{Model}
 & S1 & S2 & S3 
 & S1 & S2 & S3
 & S1 & S2 & S3 
 & S1 & S2 & S3 \\
\hline
\texttt{gemma3}  & 91.43 & 94.29 & 88.57 & 85.71 & 74.29 & 80.00 & 71.43 & 82.86 & 77.14 & 51.43 & 54.29 & 48.57 \\
\texttt{llama3}  & 82.86 & 77.14 & 74.29 & 82.86 & 85.71 & 88.57 & 71.43 & 77.14 & 77.14 & 31.43 & 31.43 & 31.43 \\
\texttt{mistral} & 80.00 & 85.71 & 85.71 & 82.86 & 54.29 & 65.71 & 74.29 & 80.00 & 74.29 & 31.43 & 28.57 & 31.43 \\
\texttt{qwen3}   & 88.57 & 80.00 & 82.86 & 60.00 & 60.00 & 68.57 & 68.57 & 71.43 & 74.29 & 43.86 & 34.29 & 34.29 \\
\hline
\texttt{gptoss} & \multicolumn{3}{c|}{82.86} & \multicolumn{3}{c|}{37.14} & \multicolumn{3}{c|}{77.14} & \multicolumn{3}{c|}{40.00} \\
\hline
\end{tabular}
\end{table}

\subsection{Insights About the Automated Assessment}
Assessing both ES and SA achieved high accuracy in most cases, with noticeable differences across rubric criteria (see~\Cref{tab:accuracy_es} and \Cref{tab:accuracy_sa}). In addition, plain-text ES are processed much faster than rich-text SA across all systems (see~\Cref{fig:sa-es-comparison}). In practice, the workflow was most reliable for structured, evidence-based rubric criteria and less reliable for criteria requiring broader contextual interpretation. Further, we discuss four practical observations below that are relevant for instructors considering AI-assisted grading.

\subsubsection{Performance Analysis (Accuracy and Time)}

The fastest configurations (often \texttt{gemma3} on S2 or S3) achieved accuracy comparable to slower alternatives on most criteria, suggesting no clear accuracy--speed trade-off (compare Tables~\ref{tab:accuracy_es} and \ref{tab:accuracy_sa}). In contrast, the performance of \texttt{gptoss} does not consistently align with small models, as its fast ES performance but weaker SA accuracy suggests an accuracy--speed trade-off. In both evaluations, \texttt{qwen3} is the slowest in all systems, except \texttt{mistral}, which was slower in SA on S2. The \texttt{qwen3} is slower because it is a reasoning model and produces larger output tokens. \texttt{gemma3} is among the fastest in all evaluations, except for \texttt{llama3} on S2 for SA. Across tasks and systems, assessing ES required less time than SA (see \Cref{fig:sa-es-comparison}). S3 was the fastest overall, while S1 was the slowest due to an older GPU. The evaluation of SA across all models and systems took longer due to the visual elements and a more complex context (see \Cref{fig:sa}). 

\subsubsection{Overall Metrics Across 10 Runs}

Based on \Cref{fig:sa-es-comparison}, ES consistently achieves higher accuracy than SA across all systems. Accuracy is also consistent across runs, as indicated by the narrow 95\% CIs around the mean for most models. \texttt{llama3} tends to lead in accuracy. Although \texttt{qwen3} has similar accuracy on ES, it is noticeably slower, particularly on S2, due to reasoning. ES also has lower response times, which is beneficial for real-time usage. \texttt{gptoss} is fastest for ES, while \texttt{gemma3} and \texttt{llama3} are the fastest for SA. In contrast, \texttt{gptoss} shows mixed performance, with competitive accuracy on some ES criteria but lower results on others, particularly in SA. These findings align with per-criterion observations where ES tasks were faster and more accurate than SA tasks. 

\subsubsection{Strict Rubrics Cause Performance Variations}

Unlike traditional evaluations of essays or open-ended questions~\cite{10.1145/3706468.3706566,Li2021,10.1007/978-3-031-36272-9_34}, where models are often given the flexibility to interpret or score along a range, our study uses strict assessment rubrics (see~\Cref{rubrics:es} and~\ref{rubrics:sa}). Most responses were graded as 0 or 1, with a single case using a 0--2 scale. In such an all-or-nothing system, partially correct answers receive 0 (which aligns with the intended grading system of the course), so small omissions or mismatches can still lead to disagreement with the ground truth score. Among different assessment criteria (see~\Cref{educational_context}), models achieved over 90\% accuracy on phishing-related tasks such as \texttt{phishing\_site\_disabled} and \texttt{phishing\_email\_address}, strongly aligning with human evaluation in those rubrics. However, performance dropped substantially for context-heavy criteria such as \texttt{sections}, where accuracy fell below 55\% (see~Tables~\ref{tab:accuracy_es} and~\ref{tab:accuracy_sa}).

\subsubsection{Contextual and Integration Challenges}

In our evaluation, models were given a predefined context of students' tasks and reports. Despite embedding that context into the prompts, the models sometimes struggled to recognize it. The results of both evaluations (\Cref{tab:accuracy_es} and~\Cref{tab:accuracy_sa}) indicate that current language models are not yet mature enough to be reliably deployed in context-dependent tasks. This highlights a key challenge for educational use: criteria such as \texttt{incident\_reports} in ES and \texttt{sections} in SA require multi-step synthesis and remain laborious for human graders. These observations highlight that while there is promise in using language models for assessment, careful consideration is needed when designing evaluation rubrics and prompts. In particular, rubrics should be based on what counts as evidence and how it is scored, which helps the model follow the provided context in its judgments rather than relying on generic reasoning.

\subsection{Implications for Teaching Practice}

Since no substantial performance trade-off was observed between larger and smaller models, smaller models demonstrate comparable or better performance with the same pipeline design. This enables a practical approach and a toolset for integrating small language models into automatic text assessment workflows, particularly in low-resource environments. 

Our study suggests that the approach is most useful for rubric criteria tied to clearly identifiable evidence in student submissions. It presents a replicable grading pipeline that uses locally run lightweight text and vision models on consumer-grade GPUs. This setup automates scoring and feedback without relying on cloud-based models. It is a cost-effective solution that reduces dependence on external services and enables offline grading. Furthermore, it demonstrates the feasibility of deploying these models in resource-constrained environments without fine-tuning, making it easier to use for instructors.

The work contributes an open-source extensible tool~\cite{supplementary-materials}, a detailed method, and lessons learned to support other instructors in adopting AI-assisted grading in diverse educational settings. The proposed method serves as a foundation for educators to replicate and adapt it for automated text assessment of student writing, including both open-ended text and rich-text responses, although our results suggest that more interpretive and context-dependent criteria still require human review. It requires minimal adjustment to accommodate various file formats, making it applicable across multiple educational and institutional contexts. Instructors can use the generated scores and feedback as an initial triage step to highlight likely score ranges and flag responses that need closer human review, especially for criteria involving document structure, synthesis, or multimodal interpretation, which they then verify, adjust, and finalize. Moreover, it encourages educators to explore AI-assisted grading not as a replacement for human assessment but as a scalable support tool that can reduce instructor workload. Also, speeding up assessment benefits students by ensuring timely feedback and enhancing consistent grading across instructors.

\subsection{Limitations and Future Work} \label{limitation}

We consider human scores to be the ground truth. There may be cases where humans made a mistake, but the model was correct, which can affect the reported accuracy. We mitigated this risk by having expert human instructors consult the lead course instructor before and during grading, and follow well-defined rubrics. The evaluation criteria were tailored to one course and need to be modified for other contexts. In addition, we observed inconsistencies in models' scoring accuracy, particularly for criteria requiring complex structure understanding or contextual interpretation. These inconsistencies may also reflect model biases, such as over-reliance on surface-level patterns or incomplete use of the provided context, as has been observed in prior work on LLM-based assessment~\cite{10.1145/3706468.3706496}. 

Future work may examine how the workflow can be adapted to other assignment types and how prompt and rubric design can better support contextual interpretation, document structure assessment, or multimodal evidence. It would also be valuable to identify which rubric dimensions are suitable for automated first-pass assessment and which continue to require closer human evaluation.

\section{Conclusions} \label{conclusion}

Prior studies on AI-assisted grading emphasized cloud-based or high-resource LLM setups, mostly overlooking locally deployable, open-source models for assessment in low-resource environments. Our study demonstrated that smaller models can be equally or even more effective than a large model for automated assessment of both text-only and rich-text documents. These accessible models can be used without fine-tuning in constrained environments, providing a practical, quick triage system for grading to support instructors. However, since our evaluation showed that some context-dependent tasks remain challenging, we emphasize that such deployment is not intended to replace human oversight. To prevent potential biases, the instructor must make the ``final call''. 

Our evaluation across systems and models tested the practical feasibility for instructors, especially those in under-resourced institutions with limited access to costly hardware or third-party services. This offers a viable, privacy-preserving path to adopting AI-assisted grading. To support transparency and reproducibility, all materials, including data, code, and evaluation templates, are publicly shared on GitLab~\cite{supplementary-materials}. The cybersecurity exercise platform we used is also available as open-source software~\cite{10.1145/3649217.3653639} -- see also \url{https://inject.muni.cz}. These resources provide a cost- and privacy-aware foundation for extending low-resource automated assessment to diverse educational domains.

\begin{acks}
This work was supported by the Ministry of Education, Youth and Sports of the Czech Republic through the e-INFRA CZ (ID:90254).
\end{acks}

\bibliographystyle{ACM-Reference-Format}
\bibliography{ref}
\end{document}